\documentclass[aps,pre,showpacs,preprint]{revtex4-1}

\usepackage{amsmath} 
\usepackage{graphicx}
\usepackage{amssymb} 
\usepackage{chemarr} 
\usepackage{wrapfig}

\begin{document}
\title{Network analysis of nanoscale energy conversion processes}

\author{Mario Einax}
\email{einaxm@biust.ac.bw}
\affiliation{Department of Physics and Astronomy, Botswana International University of Science and Technology, Palapye, Botswana}

\date{\today}

\begin{abstract}
Energy conversion in nanosized devices is studied in the framework of state-space models.
We use a network representation of the underlying master equation to describe the
dynamics by a graph. Particular segments of this network represent input and
output processes that provide a way to introduce a coupling to several heat reservoirs and particle reservoirs.
In addition, the network representation scheme allows one to decompose the stationary dynamics
as cycles. The cycle analysis is a convenient tool for analyse models of machine operations,
which are characterized by different nanoscale energy conversion processes. By
introducing the cycle affinity, we are able to calculate the zero-current limit.
The zero-current limit can be mapped to the zero-affinity limit in a network representation scheme.
For example, for systems with competing external driving forces the open-circuit voltage can be determined
by setting the cycle affinity zero. This framework is used to derive
open-circuit voltage with respect to microscopic material energetics and different
coupling to particle and temperature reservoirs.
\end{abstract}

\maketitle

\section{Introduction}
\label{sec:intro}
Energy conversion in nanosized device architectures, where competing external driving forces are involved, is the
focus of intensive current research \cite{Seifert:2012}.
Prominent examples are organic photovoltaic cells \cite{Rutten/etal:2009,Markvart:2007,Kirchartz/Rau:2008,Einax/etal:2011}, thermoelectric devices \cite{Jiang/etal:2012,Arrachea/etal:2014,Entin-Wohlman/etal:2015}, quantum dots \cite{Jordan/etal:2013} or peristaltic pumps \cite{Einax/etal:2010a,Dierl/etal:2014}. Energy converters operate inherently under nonequilibrium conditions.
The quest to improve the energy harvesting process of low-dimensional energy conversion materials with several external driving forces
is naturally linked to a deeper understanding of underlying carrier transport. For example, in thermoelectric transport systems electrical voltages can be generated by temperature differences. At the macroscopic level, the general relations between heat and particle currents and their responses to temperature differences and applied voltages are given in terms of linear irreversible thermodynamics, however, an adequate description of
low-dimensional nanoscale energy conversion devices requires microscopic approaches to model the coupling to external reservoirs and driving forces.

Modeling of carrier transport of nanosized device architectures is often done in terms of a state-space approach.
In such an approach,  the kinetics is described by a master equation.
In such state-space models the processes underlying the device operation are given as transitions
between microscopic system states. Our approach is formulated in the framework of network (graph)
theory \cite{Schnakenberg:1976}. In the graph theory approach, the master equation is represented
by a graph \cite{Seifert:2011,Einax/Nitzan:2014,Polettini/etal:2016}
that consists of a set of edges and nodes. The nodes correspond to states, while the edges represent transitions between states.
When focusing on the steady state operation of nanosized device architectures,
the graph theory approach can be used to construct a decomposition of the graph topology into cycles representing the steady state dynamics of the device.
Each cycle is characterized by a cycle affinity.
In this scheme, it is straightforward to include external parameters,
which maintaining the system in a non-equilibrium state.
For efficient device operation, it is of particular relevance to
understand the interplay between different input (driving) and output (motion against load) processes.
Driving and load processes can be associated to particular segments in a cyclic network path.

A load tends to reduce the current generated by driving force until the current becomes zero.
In many applications, such as photovoltaic cells or thermoelectric devices, the stopping point of the system current is denoted as
the open-circuit (OC) voltage. Of crucial importance for understanding energy conversion processes in those systems is how the
open-circuit voltage is connected to both the microscopic energetics and the couplings to the environment.
Recently, it was shown that for both organic photovoltaic device setups and thermoelectric energy harvester
those relations can be derived by using the cycle analysis approach \cite{Einax/Nitzan:2016}.
In this approach, the analytical expression for the open-circuit voltage
follows by setting the cycle affinity of a basic transport cycle to zero.
The advantage of using such cycle analysis is that it is easily
generalizable to any model that can be represented by a
network of states and rates.

For nanosized device architectures, the details of the interplay between system and its coupling
to the environment are essential in modelling the energy conversion. Because of the discrete nature
of the energetics (energy level) in such devices, the reservoir influence on the transition rates between states
is determined by the specific nature of the system-reservoir coupling leading to both particle and energy
exchange between system and environment. In many artificial nanosized devices, the transition rates are determined by a single heat bath at a given temperature, however, different transition rates can be linked with a specific single heat bath having its own temperature \cite{Jiang/etal:2012,Entin-Wohlman/etal:2015,Esposito/etal:2009,Esposito/Lindenberg:2009,Einax:2014}.
In principle, there is no compelling need to have only one heat bath involved in transition processes.
The seminal work of Trimper \cite{Trimper:2006} is a prominent example. In that paper the author discussed a simple spin-flip model with Glauber dynamics under the presence of two heat reservoirs with different temperatures, which are coupled separate to the two possible flip processes.
Another examples is given in Refs. \cite{Craven/Nitzan:2017,Craven/etal:2018}. The authors have considered electron transfer dynamics in a thermally heterogeneous environment, and their approach is used to model carrier transport between two sites of different local temperatures.

Here, we address a similar question for a simple nanosized energy conversion device. The system is coupled to a single heat reservoir at temperature $T_M$, while the particle injection and ejection rates at the system boundary are coupled separately to heat reservoirs
which are characterized by their respective temperatures. For example the particle injection from an electrode is controlled by a head bath at temperature $T'$, while the inverse process is controlled by the heat bath at temperature $T_M$. We apply the cycle analysis method to calculate
the open-circuit voltage in the limit of vanishing currents, which corresponds to the maximal useful work done by the device.

\section{Model}
\label{sec:model}
\begin{figure}[h!]
 \centering
 \includegraphics[width=0.45\columnwidth]{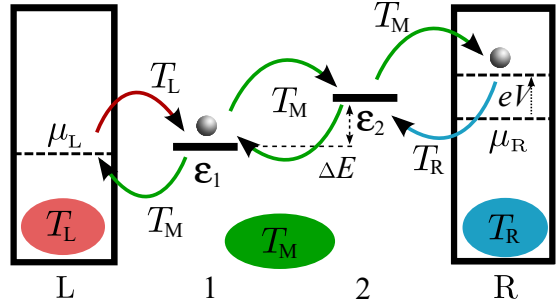}
 \caption{Schematic diagram of the energy conversion device. The contacts (electrodes) on the left and the right are characterized by their
 respective electrochemical potentials $\mu_L$ and $\mu_R$, temperatures $T_L$ and $T_R$. The system is controlled by temperature $T_M$.
  The particle injection from the left and right contact is determined by rates which are controlled solely by the temperature $T_L$ (red curved arrow) and $T_L$ (blue curved arrow). The green curved arrows represent transitions controlled by temperature $T_M$.}
 \label{fig:fig1}
\end{figure}

Our discussion is based on the assumption that a suitable (coarse-grained) state-space model can be constructed for the energy conversion process of interest. For simplicity we consider a simple nanosized device, which is illustrated in Figure~\ref{fig:fig1}.
The nanodevice comprising a two-level system situated between two external
contacts, $L$ and $R$  and a given energy landscape characterized by the site energy levels $\varepsilon_k$ with $k = 1,2$.
The gap energy $\Delta E= \varepsilon_2-\varepsilon_1$ denotes the energy difference between $\varepsilon_2$ and $\varepsilon_1$
While the system is held at temperature $T_M$, the two contacts are kept at different temperature $T_L$ and $T_R$ respectively.
Neglecting particle-particle interactions, the charge carrier transport is characterized by three states:
two-level system: $0$ (vacant), $1$ (charge carrier at site $1$), and $2$ (charge carrier at site $2$).
The two metallic left and right contacts are realized as two electrodes (particle reservoirs) and characterized by their chemical potentials $\mu_L$, and $\mu_R$, respectively. The electrochemical potential difference corresponds
to a bias voltage $V=(\mu_R-\mu_L)/|e|$, where $|e|$ is the electron charge.
The particle exchange between the left electrode $L$ and site $1$ involve two separate heat reservoirs at different temperatures $T_L$ and $T_M$, i.e.,
site $1$ is coupled to two heat baths. While the injection of a particle into the system on site $1$ is controlled solely by the heat bath at temperature $T_L$, the ejection of a particle to the left contact is controlled solely by the system temperature $T_M$. Analogously, the ejection of particles to the right contact is coupled solely to a heat reservoir at temperature $T_M$, while the injection from the right contact is coupled solely to a heat bath at temperature $T_R$. The hopping of a particle within the system (between site $1$ and $2$) is is controlled by the temperature $T_M$.

The probabilities to find the system in a state $i=0,1,2$ are given by $P_i (t)$ fulfilling normalization
$\sum_i P_i(t)=1$ at all times. Thus, the system dynamics is modeled by a master equation
\begin{align}
\label{eq:master_equation_current}
\frac{d P_i}{dt} &= \sum_{j=0}^{2} J_{ij} (t)
\end{align}
accounting for the time evolution of the probabilities $P_i (t)$, where the (net) link probability current
from state $j$ to state $i$ reads $J_{ij} (t)= k_{ij} P_j (t) - k_{ji} P_i (t)$ \cite{Sylvester-Hvid/etal:2014,Einax/etal:2010b,Einax/etal:2013}.
The steady state solution is given by the condition $\frac{d P_i}{dt}=0$ for all $i$.

\begin{figure}[h!]
 \centering
 \includegraphics[width=0.42\columnwidth]{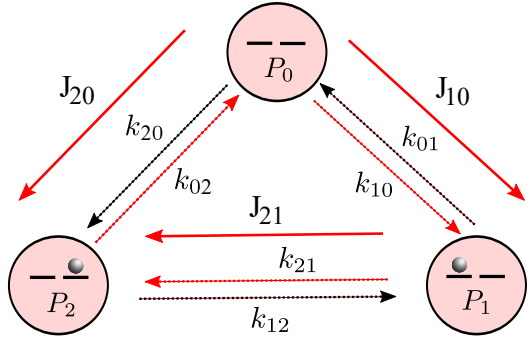}
 \caption{Network representation of the underlying master equation associated with the three accessible states.
  The graph is composed of 3 vertices (shown as circles). The interconnected vertices represent the probabilities
  $P_j$ to find the system in a state $j$ ($j=0,1,2$) and the edges connecting some pairs of vertices.
  The edges are drawn as arrows that indicate transitions with rate $k_{ij} = k_{i\leftarrow j}$
  from a state (vertex) $j$ to $i$.}
 \label{fig:fig2}
\end{figure}

The rate $k_{ij}=k_{i \leftarrow j}$ controls the transition from a state $j$ to a state $i$.
The transition rates between states $j=0,1,2$ are determined by the state energy $E_0=0$, $E_1=\varepsilon_1$, and $E_2=\varepsilon_2$,
bias voltage $V=(\mu_R-\mu_L)/|e|$, and the local temperatures $T_L$, $T_R$, and $T_M$.
Figure~\ref{fig:fig2} shows the connectivity network topology for the introduced energy conversion device,
which is a simple cyclical graph that represent the following path
\begin{align}
\label{eq:fb_path}
0 & \xrightleftharpoons[k_{01}]{k_{10}} 1 \xrightleftharpoons[k_{12}]{k_{21}} 2 \xrightleftharpoons[k_{20}]{k_{02}} 0
\end{align}
in the underlying state-space.

The rates $k_{12}$ and $k_{21}$ are associated with a single heat bath at temperature $T_M$ satisfying the standard
detailed balance conditions
\begin{align}
\label{eq:system_condition}
\frac{k_{21}}{k_{12}} = e^{-\beta_M (E_2-E_1)}\, ,
\end{align}
where $\beta=1/k_{\rm\scriptscriptstyle B} T_M$ is the inverse thermal energy with
$k_{\rm\scriptscriptstyle B}$ being Boltzmann constant. Following the derivation in Ref.~\cite{Trimper:2006},
the transitions between state $0$ and state $1$ ($0 \xrightleftharpoons[k_{01}]{k_{10}} 1$) and between state $2$ and state $0$
($2 \xrightleftharpoons[k_{20}]{k_{02}} 0$) are coupled in each case to two head baths including
the particle exchange of the system with the left and right contact, respectively.
Particles are injected or ejected from the two reservoir sites with rates that still fulfil the condition of detailed balance \cite{note1}
with respect to the generalized grand-canonical ensembles associated with $\mu_L$ and $\mu_R$ and effective temperatures
\begin{align}
\label{eq:T_eff}
T_L^e=\frac{2T_L T_M}{T_L+T_M}\quad \textrm{and} \quad T_R^e=\frac{2T_R T_M}{T_R+T_M} \, .
\end{align}
The corresponding rates satisfy the relations
\begin{align}
\label{eq:left_condition}
\frac{k_{10}}{k_{01}} &= e^{-\beta_L^e (E_1-\mu_L)}\, ,\\
\label{eq:right_condition}
\frac{k_{20}}{k_{02}} &= e^{-\beta_R^e (E_2-\mu_R)}\, .
\end{align}
with $\beta_L^e = (\beta_L +\beta_M)/2$ and $\beta_R^e = (\beta_R +\beta_M)/2$, respectively. Thus, the effective temperature is given by the harmonic average of the two temperatures triggering the particle exchange between system and environment.

\section{Cycle analysis}
\label{sec:calculation}
The steady state operation of the nanodevice introduced in Fig. \ref{fig:fig1}
can be investigated in terms of a cycle analysis of the cyclic graph
shown in Fig.~\ref{fig:fig2}.
The central quantity is the cycle affinity $\mathcal{A}$, which is defined by \cite{Schnakenberg:1976}
\begin{align}
\label{eq:cycle_affinity}
\mathcal{A} &= -\ln \mathcal{K} \, ,
\end{align}
where $\mathcal{K}$ is the ratio between products of forward and backward rates of the closed path
\begin{align}
\label{eq:ratio_cycle_orig}
\mathcal{K} &=\frac{k_{1,0}k_{2,1} k_{0,2} }{k_{0,1} k_{1,2} k_{2,0}  }  \, .
\end{align}
Note that passing through the cycle in clockwise direction transports a particle from the left to the right contact electrode.
Using Eq.~(\ref{eq:system_condition}))and Eqs.~(\ref{eq:left_condition}-\ref{eq:right_condition}) in Eq.~(\ref{eq:ratio_cycle_orig}) yields
\begin{align}
\label{eq:ratio_cycle_C1}
\mathcal{K} &  =
e^{-\beta_{L}^e (\varepsilon_1-\mu_{\rm L})} e^{-\beta_M (\varepsilon_2-\varepsilon_1)} e^{\beta_R^e (\varepsilon_2-\mu_{R})}
\equiv e^{-\mathcal{A}} \, ,
\end{align}
and the affinity of the cycle is given by
\begin{align}
\label{eq:affinity}
\mathcal{A} &= \beta_{L}^e (\varepsilon_1-\mu_{\rm L}) + \beta_M (\varepsilon_2-\varepsilon_1) - \beta_R^e (\varepsilon_2-\mu_{R}) \, .
\end{align}
Setting $\mathcal{A}=0$, i.e. $\mathcal{K}=1$, we arrive at the stopping condition
\begin{align}
\label{eq:stopping_condition}
\frac{1}{2} \left( \beta_R \mu_R - \beta_L \mu_L \right) + \frac{\beta_M}{2} \left( \mu_R - \mu_L \right) &=
- \frac{\beta_M}{2} \left( \varepsilon_2 - \varepsilon_1 \right) + \frac{\varepsilon_2}{2}\beta_R - \frac{\varepsilon_1}{2}\beta_L \, ,
\end{align}
where all currents through the system vanish.

\section{Results and discussion}
\label{sec:results}
We consider now the special case $\beta_L=\beta_R=\beta=1/k_BT$. Inserting $|e| V_{\rm OC}=\mu_R-\mu_L$ and $\Delta E= \varepsilon_2-\varepsilon_1$
in Eq.~(\ref{eq:stopping_condition}), it follows that
\begin{align}
\label{eq:condition_Voc}
 \left( 1 + \frac{T}{T_M} \right) |e| V_{\rm OC} &= \left( 1 - \frac{T}{T_M} \right)\Delta E \, .
\end{align}
By introducing the Carnot efficiency $\eta_C = 1-T/T_M$, we arrive at the central result
\begin{align}
\label{eq:Voc_new}
|e| V_{\rm OC} & = \frac{\eta_C}{2-\eta_C} \Delta E\, .
\end{align}
If we replace in the detailed balance relations (\ref{eq:left_condition}) and (\ref{eq:right_condition}) the effective temperatures by the temperature $T$, i.e., the injection and ejection is controlled by a single heat bath at temperature $T$, we arrive at
\begin{align}
\label{eq:Voc_previous}
|e| V_{\rm OC} &= \eta_C \Delta E \, .
\end{align}
\begin{figure}[h!]
 \centering
 \includegraphics[width=0.65\columnwidth]{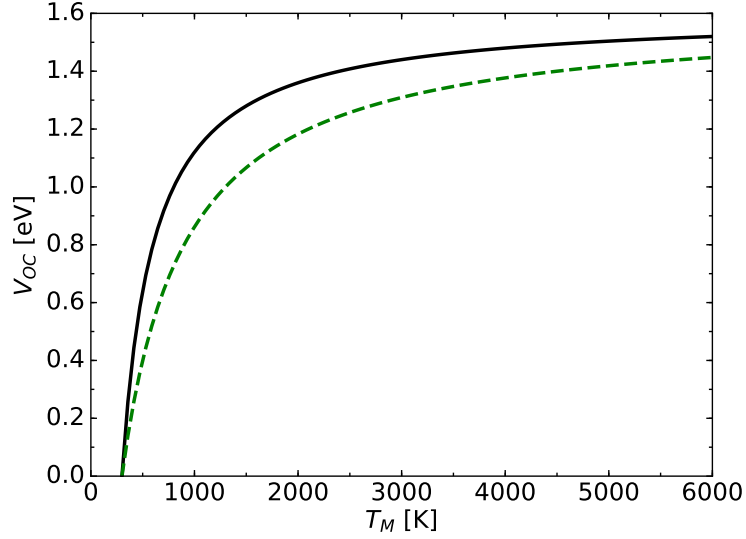}
 \caption{Open-circuit voltage as function of the system temperature $T_M$ for $T=300$K and $\Delta E=1.6$eV.
  The solid line corresponds to Eq.~(\ref{eq:Voc_previous}), while the dashed line visualizes Eq.~(\ref{eq:Voc_new}).}
 \label{fig:fig3}
\end{figure}
In the limits $\eta_C=0$ ($T=T_M$) and $\eta_C=1$ (finite $T$ and $T_M \rightarrow \infty$), Eqs.~(\ref{eq:Voc_new}) and (\ref{eq:Voc_previous}) approach each other.

To illustrate the different performance resulting form Eqs.~(\ref{eq:Voc_new}) and (\ref{eq:Voc_previous}) the following set of parameters are used:
$T=300$K and $\Delta E=1.6$eV. The variations of the open-circuit voltage as function of $T_M$ are plotted Fig. \ref{fig:fig3}.
This suggests that, if the injection and ejection rates of particles are conditioned by different temperatures, the performance of this nanodevice setup is less efficient compared to a nanodevice setup, in which the injection and ejection rates of particles are conditioned by only a single temperature.

Now we assume that we have a system temperature $T_M$ which is different from the temperature $T_L$ at the left contact and the temperature $T_R$ at the right contact and $\beta_L \neq \beta_R$ with $\beta_L=1/(k_B T_L)$ and $\beta_R=1/(k_B T_R)$, i.e., both contacts are kept on different temperature. We also assume that $\mu_L=\varepsilon_F-V_{\rm OC}/2$ and $\mu_R=\varepsilon_F+V_{\rm OC}/2$ and $\varepsilon_F$ being the Fermi energy, which is the same for both metallic contacts. In that case
we obtain, when starting from Eq.~(\ref{eq:stopping_condition}), the following expression for the stopping voltage
\begin{align}
\label{eq:Voc_three_new}
|e| V_{\rm OC} & = 2 \left[ \varepsilon_2 \frac{ (\beta_R-\beta_M)}{\beta_L+\beta_R+2\beta_M} - \varepsilon_1 \frac{ (\beta_L-\beta_M)}{\beta_L+\beta_R+2\beta_M} - \varepsilon_F \frac{ (\beta_R-\beta_L)}{\beta_L+\beta_R+2\beta_M} \right]\, .
\end{align}

\section{Conclusion}
\label{sec:conclusion}
In summary, we have considered energy conversion processes under the influence of an effective temperature. The effective temperature controls the injection and ejection of particles and appears via the coupling to individual heat reservoirs of the injection and ejection process, i.e., the particle exchange between system and electrodes is coupled to more than one heat bath.
Finally, our work has shown that, in nanoscale devices, where the particle exchange between system and electrodes is
coupled to more than one heat bath, the particle flow and the energy conversion reduce significantly and its consequence is
manifested in a reduced stopping (OC) voltage.
Analytical expressions for the open-circuit were derived from the affinity associated with a network scheme of
the underlying rate processes. Finally, the approach presented in this paper can also be used
to study the heating by cooling process in a three-terminal thermoelectric device design, where injection and ejection
processes of electrons are coupled differently to temperature reservoirs, i.e.,  the injection process is solely determined by the temperature of the metallic contact, while the ejection process is solely triggered by the systems temperature.

\section*{Acknowledgments}
M.E. thanks A. Nitzan for illuminating discussions on the coupling of multiple heat reservoirs in the context of charge transfer processes at the Tel Aviv University and FU Berlin. M.E. also acknowledges funding by a Research Initiation Grant at BIUST (Grant No. R00103).

\end{document}